\documentclass[aps,prl,twocolumn, superscriptaddress]{revtex4}
\usepackage[dvips]{graphicx}
\usepackage{amssymb}
\usepackage{amsmath}
\usepackage{color}
\usepackage{gensymb}
\begin{document}

\title{Competing charge density waves probed by non-linear transport and noise in the second and third Landau levels }

\author{K. Bennaceur}
\affiliation{D\'{e}partement de Physique et Institut Quantique, Universit\'{e} de Sherbrooke, Sherbrooke, Qu\'ebec, J1K 2R1, Canada}
\affiliation{Department of Physics, Amrita Vishwa Vidyapeetham, Amritapuri, India}
\affiliation{Department of Physics, McGill University, Montr\'eal, Qu\'ebec, H3A 2T8, Canada}

\author{C. Lupien}
\affiliation{D\'{e}partement de Physique et Institut Quantique, Universit\'{e} de Sherbrooke, Sherbrooke, Qu\'ebec, J1K 2R1, Canada}

\author{B. Reulet}
\affiliation{D\'{e}partement de Physique et Institut Quantique, Universit\'{e} de Sherbrooke, Sherbrooke, Qu\'ebec, J1K 2R1, Canada}

\author{G. Gervais}
\affiliation{Department of Physics, McGill University, Montr\'eal, Qu\'ebec, H3A 2T8, Canada}

\author{L. N. Pfeiffer}
\affiliation{Department of Electrical Engineering, Princeton University, Princeton NJ 08544 USA}

\author{K. W. West}
\affiliation{Department of Electrical Engineering, Princeton University, Princeton NJ 08544 USA}

\date{\today}
\begin{abstract}

Charge density waves (CDW) in the second and third Landau levels (LL) are investigated by both non-linear electronic transport and noise. The use of a Corbino geometry ensures that only bulk properties are probed, with no contribution from edge states. Sliding transport of CDWs is revealed by narrow band noise in re-entrant quantum Hall states R2a and R2c of the second LL as well as in pinned CDWs of the third LL.  Competition between various phases - stripe, pinned CDW or fractional quantum Hall liquid - in both LL are clearly revealed by combining noise data with maps of conductivity versus magnetic field and bias voltage.

\end{abstract}


\maketitle

Spontaneous charge ordering is one of the many intriguing quantum phenomena occurring in a 2D electron gas (2DEG) under a magnetic field. In the first (N=0) Landau level (LL), the short range attractive part of Coulomb interaction is known to be responsible for the condensation of quasiparticles  into an incompressible Laughlin liquid \cite{Laughlin} hosting a fractional quantum Hall effect (FQHE) and carrying fractional charges. In higher LLs ($N\geqslant 2$), however, the situation is markedly different as the combination of short-range attractive with long-range repulsive Coulomb interaction leads to electronic phases forming charge density waves (CDWs). At half filling factor of these LLs, $\nu^* = \nu - [\nu] = 1/2$, with $[\nu]$ the integer part of $\nu = n_{s}h/eB$ (where $ n_{s}$ is the electron density and $B$ the magnetic field) , an alternation between stripes of filling $\nu^* = 1$ and $\nu^* = 0$ with a spatial period of the order of the cyclotron length leads to a ground state with broken symmetry believed to be a smectic liquid crystal. At finite temperature, a Berezinskii-Kosterlitz-Thouless (BKT) transition is predicted \cite{Fradkin, Orion} between a smectic and a nematic phase, followed by a melting at higher energy restoring the isotropic phase. Several observations of anisotropic transport supporting these theories have been reported \cite{9/2stripe1,9/2stripe2,N2Yacobi, Lloyd2008}. 
At lower partial filling factor $\nu^* \approx M/(3N)$, Hartree-Fock calculations \cite{Koulakov1995, Fogler2002} also predict a pinned bubble crystal with $M$ electrons (or holes) per bubble which was observed with microwave conductivity measurements \cite{Lloyd2002}. In between such phases, at $\nu = 1/4$ and $\nu = 3/4$ a pinned CDW phase has been observed \cite{WCDepin} and, albeit still debated, could be composed of stripe segments with a depinning transition to a nematic electron liquid crystal at finite bias voltage \cite{N2Yacobi}.

In the second ($N=1$) Landau level (SLL), CDW and FQHE are known to compete, and a re-entrant quantum Hall effect (RIQHE) is often observed amongst various FQHE states when the electron mobility is sufficiently high \cite{reentrant1}. The RIQHE is characterized by a plateau in the Hall resistance at given non integer values of partial filling factor  $\nu^*$.  Although there are experimental evidences that RIQHE states are CDWs, their nature is still being highly debated \cite{NMR1,NMRN2,josh}. At first sight, the absence of transport anisotropy suggested charge ordering in the form of bubble phases, however recent experiments involving resistively detected nuclear magnetic resonances (RDNMR)\cite{NMRN2} show that RIQHE phases near $\nu^*=1/2$ (R2b and R2c) host polarized and unpolarized spin regions, suggestive of a  stripe phase. In contrast, another RDNMR experiment \cite{NMR1} focusing on RIQHE phases at $\nu^* < 1/3$ (R2a and R2d), suggests that these are more likely bubble phases. Furthermore, other recent works show that in spite of a competition, CDW and FQHE phases could even coexist at a same $\nu^*$ \cite{phaseComp1, phaseComp2}.

Most of previous transport investigations of CDW  phases have been performed in Van der Pauw (VdP) or Hall bar geometries where both bulk and edge transport contribute.  With the exception of Refs.\cite{CorbiDepin, Schmidt2015, Fu2017, Wang2017}, and to our knowledge,  only the resistivity  and {\it not} the conductivity  has been mapped out as function of DC bias in the SLL \cite{josh} and in the TLL \cite{N2Yacobi}. This is precisely the purpose of this work to probe bulk electronic transport properties in both the SLL and TLL of ultra-high mobility Corbino-shaped 2DEG samples. The differential conductance $G=\partial{I}/\partial{V}$ as well as the current noise spectral density $S_{II}$ was measured as a function of DC bias voltage $V_{DC}$ and magnetic field in samples with different ring sizes. This allowed us to draw several conclusions: (i) transport involves sliding CDWs, suggesting 
a stripe-like order CDW such as a nematic electron liquid crystal in the RIQHE of the SLL and away from half filling in the TLL ; (ii) a bubble-phase scenario for three RIQHE states in the SLL is unlikely; (iii) FQHE and CDWs can coexist in the SLL. 

\begin{figure}[!h]
\includegraphics[width=8cm]{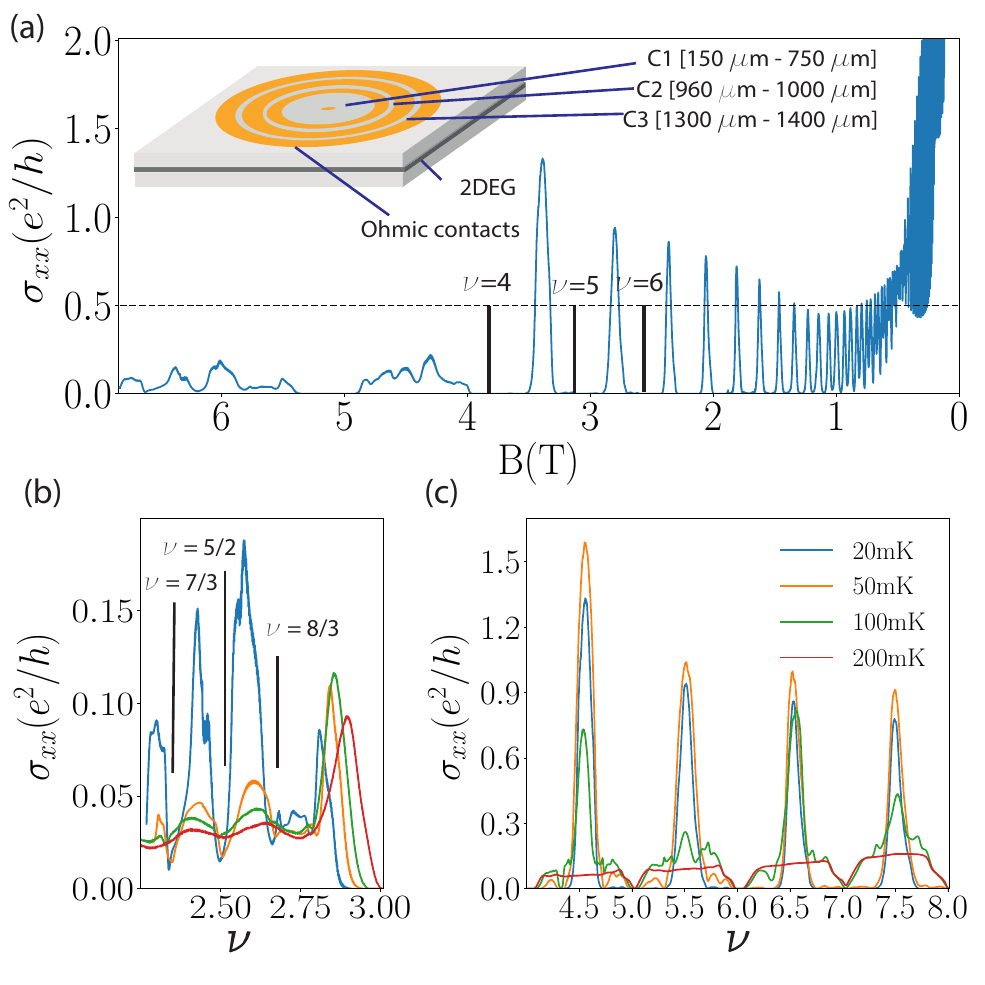}
\caption{(a) Conductivity $\sigma_{xx}$ \textit{vs.} magnetic field $B$ for sample C1. Inset: Schematics of the samples. (b) and (c): $\sigma_{xx}$ of sample C1 at $20\,\mathrm{mK}$ (blue curve), $50\,\mathrm{mK}$ (orange curve), $100\,\mathrm{mK}$ (green curve), and $200\,\mathrm{mK}$ (red curve) in the SLL lower spin branch (b) and in the $N=2$ and $N=3$ LL (c). }
\label{C1carac}
\end{figure}

Three samples in concentric Corbino geometries have been fabricated on the same 2DEG, and they are depicted in the inset of Fig. \ref{C1carac}(a). The 2DEG is a quantum well of GaAs/AlGaAs located $200\, \mathrm{nm}$ underneath the surface, with two Si doping layers $100\, \mathrm{nm}$ above and below the 2DEG, yielding a charge carrier density of $n_s = 3.8\, \times 10^{11} \mathrm{cm}^{-2}$. Details pertaining to the fabrication, cooldown procedure, and noise measurements are provided in \cite{suma}. Except when noted, all measurements have been performed in a dilution refrigerator with a base temperature of $7 \, \mathrm{mK}$.  From the temperature dependence of the conductance in FQHE states,  we estimate the electron temperature to saturate around $20 \, \mathrm{mK}$. The electron mobility in the Corbino geometry cannot be determined precisely because of the inherent two-point contact measurement, however the onset of Shubnikov de Haas (SdH) oscillations at very low fields, $B\simeq 0.03\, \mathrm{T}$, is highly suggestive of an extremely high mobility. In addition, we measured a mobility of $\mu = 2.5\times 10^7 \,\mathrm{cm}^2 \,\mathrm{V}^{-1}\, \mathrm{s}^{-1}$ on a Hall bar sample cut from the same wafer (on a different cooldown), also suggestive of the extremely high quality of the Corbino samples used. 

The zero-bias conductivity  $\sigma_{xx}$ \textit{versus} magnetic field is shown in Fig. \ref{C1carac}(a) for sample C1. The transition between SdH and integer quantum Hall regime occurs at $B\sim1 \, \mathrm{T}$. There, the conductivity $\sigma_{xx}$ maxima are very close to $0.5\,e^2/h$, as expected \cite{SigMaxCond, SigMaxCond2}, and vary by less than $\sim10\%$ for temperature between $20 \, \mathrm{mK}$ and $200 \, \mathrm{mK}$ in both spin-resolved branches. At higher magnetic fields in Landau levels $N<5$, $\sigma_{xx}$ departs from $0.5\,e^2/h$. As shown in Fig. \ref{C1carac}(c), the conductivity maxima vary non monotonically with temperature and increase from $20 \, \mathrm{mK}$ to $50 \, \mathrm{mK}$, reaching $\sim 1.7\,e^2/h$ at $\nu=9/2$, and then decrease with temperature increasing. Finally, we note that the temperature dependence of the spin down phases $\nu=9/2$ and $\nu=13/2$ are more robust than for the spin up $\nu=11/2$ and $\nu=15/2$, as has been observed in the stripe phase in the VdP geometry \cite{9/2stripe1, 9/2stripe2}.

\begin{figure}[!h]
\includegraphics[width=9cm]{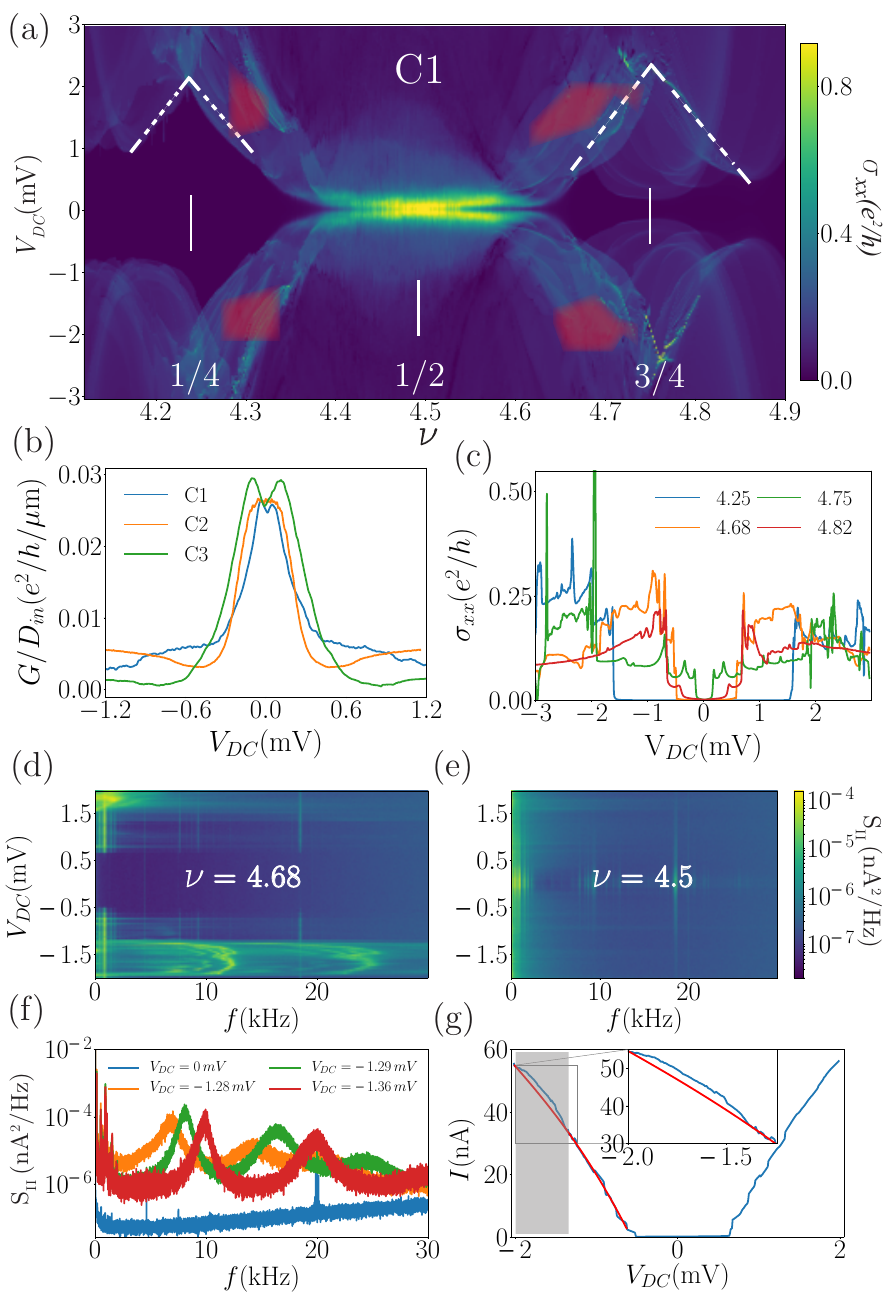}
\caption{(a) $\sigma_{xx}$ in the TLL \textit{vs.} $V_{DC}$ and $\nu$ in C1. Narrow band noise was observed in red overlay regions.  (b) $G/D_{in}$ ($G$ being the conductance and $D_{in}$ the inner contact diameter) \textit{vs.}  $V_{DC}$ at $\nu = 9/2$ in the three samples. (c) $\sigma_{xx}$ \textit{vs.} $V_{DC}$ at $\nu = 4.25,\,4.68,\,4.75,\,4.82$ in C1.  (d,e) Noise spectral density $S_{II}(f)$ \textit{vs.} frequency $f$ at $\nu = 4.68$ and 4.5. (f) $S_{II}(f)$ \textit{vs.} frequency $f$ at $\nu = 4.68$ for $V_{DC} = 0 \mathrm{mV}, -1.28 \mathrm{mV}, -1.29\mathrm{mV}, -1.36 \mathrm{mV}$. (g) Blue: DC current $I$ \textit{vs.} $V_{DC}$ at $\nu =4.68$ with a zoom on narrow band noise region in central inset. Red: cubic interpolation of $I(V_{DC})$. The greyed zone is where narrow band noise occurs.}
\label{QHE1}
\end{figure}

Our measurements in the TLL are reported in Fig. \ref{QHE1}.  Fig. \ref{QHE1}(a) shows a color intensity plot of $\sigma_{xx}$ {\it versus} DC bias voltage $V_{DC}$ and filling factor $\nu$ in the lower spin branch of the TLL ($4<\nu<5$) for sample C1. The non-linear transport measurements reveal important features that are absent at $V_{DC}=0$. We first focus on the central region corresponding to the stripe phase around $\nu=9/2$. Corbino samples have rotational invariance,  however the anisotropy of the stripe phase appears as follows. The conductivity in the Corbino geometry is given by $\sigma_{xx} = \frac{G}{2\pi} \ln(D_{out}/D_{in})$, where $G$ is the differential conductance and $D_{in}$ ($D_{out}$) is the inner (outer) ring diameter. In an isotropic phase, $\sigma_{xx}$ should not depend on the sample size. Yet, we observe that in the (presumably anisotropic) stripe phase, the conductivity ratios of the three different samples are far from the expected value of 1: at 20~mK and $V_{DC} = 0\,V$, $\sigma_{xx}^{C1}/\sigma_{xx}^{C2} = 2.6$ and  $\sigma_{xx}^{C1}/\sigma_{xx}^{C3} = 6.8$. This phase is known to have a preferred direction (the so-called easy axis), and as a consequence we rather expect the conductance to be proportional to the number of stripes linking the two contacts, {\it i.e.} in our case to $D_{in}$.  This is what is observed here: as shown in Fig. \ref{QHE1}(b), the ratio of conductance $G$ at $V_{DC} = 0\,V$ over $D_{in}$ is $G/D_{in} = 0.026\, e^2/h \, \mu \mathrm{m}^{-1}$, as extracted from a linear fit of $G$ {\it versus} $D_{in}$, see \cite{suma}. In contrast, at 200~mK where the stripe phase is expected to have melted, the conductivity ratios are found to be  $\sigma_{xx}^{C1}/\sigma_{xx}^{C2} \approx 0.7$, and much closer to the expected value of one. Finally, we observe three regime of conductance for samples C1 and C3 (see Fig. \ref{QHE1}(b)) with increasing bias voltage : one that increases with $V_{DC}$, a second where it decreases,  and a third nearly constant. This is consistent with a smooth BKT transition between a smectic to a nematic phase and then to an isotropic phase, as observed recently in Ref.\cite{Manfra}

At filling factors deviating from $\nu=9/2$, the differential conductance is observed to vary abruptly at non-zero DC bias forming diamond shape regions, see dashed lines in Fig. \ref{QHE1}(a). In addition, a pronounced hysteresis upon voltage sweep direction is also observed. This is consistent with the de-pinning transition typical of a pinned CDW \cite{WCDepin,CorbiDepin}. The diamonds are located at filling factors in the vicinity of $\nu^*  = 1/4$ and $\nu^*  = 3/4$, as was previously reported \cite{WCDepin,CorbiDepin, N2Yacobi}. We note the situation is however more complex for $\nu^*  = 3/4$ where  $\sigma_{xx}$ has several jumps which could be different de-pinning transitions occurring at different threshold bias voltages, as exemplified by the green curve of Fig. \ref{QHE1}(c).

The current noise spectral density $S_{II}(f)$ as a function of DC bias voltage $V_{DC}$ was measured in a DC to 65 kHz bandwidth. Examples of such spectra are presented in Fig. \ref{QHE1}(f) for fixed DC bias voltage. The voltage dependence of the spectra are shown as color intensity plots in Fig. \ref{QHE1}(d,e) for sample C1 at two different filling factors in the TLL. In certain ranges of DC bias voltage, the noise spectra show peaks at finite frequencies, often followed by several harmonics as shown for $\nu = 4.68$ in Fig. \ref{QHE1}(f). This narrow band noise (NBN) occurs at the boundary separating stripes and pinned CDWs around $\nu^* \approx 0.3$ and $\nu^* \approx 0.7$ and often at bias voltage significantly above depinning threshold voltage (see regions in red overlay in Fig. \ref{QHE1}(a)). NBN was reported before in \cite{SCDW} for  $\nu = 4 + 1/4$ and $6 + 1/4$ however there was no clear understanding of these phenomena. The NBN observed here is reminiscent of  ``washboard noise'' associated with sliding mode conductivity of CDWs \cite{1slide}. In this scenario, the current carried by a sliding CDW is given by $I_{CDW}= e f_0  \lambda n_{c}$, where $f_0 = v_d/ \lambda$ is the fundamental frequency (here of the order of $10$ kHz), $\lambda$ is the periodicity of the CDW, $v_d$ the drift velocity and $n_c$ is the density of electrons condensed in the sliding CDW.  As there is already current flowing in the sample when NBN occurs it is difficult to measure precisely $I_{CDW}$ as a function of bias voltage. However,  by interpolating the DC current versus bias voltage $I(V_{DC})$ (blue curve in Fig. \ref{QHE1}(g)) with a cubic polynomial (red curve in Fig. \ref{QHE1}(g)) and subtracting it to $I(V_{DC})$ (see \cite{suma}), a current increase is observed. The voltage range in which an increase of the DC current is observed coincides with that where NBN is observed (shaded region in Fig. \ref{QHE1}(g)).  At $\nu = 4.68$, we find  $I_{CDW}/f_0 = 2.46 \times 10^{-13}\, \mathrm{A\,s}$.  Assuming further a CDW periodicity on the order of the cyclotron radius (\cite{Koulakov1995}) $l_c=\sqrt{N+1}\, l_B \sim 31.8 \,\mathrm{nm}$, ($l_B$ being the magnetic length), an estimate for the electron density in the CDW is found to be  $n_{c} \sim 4.8\times 10^{9} \, \mathrm{cm}^{-2}$. The velocity of the sliding CDW can also be estimated, $v_d \sim 380\, \mu \mathrm{m\,s^{-1}}$. Given the electron density $n_s = 3.8 \times 10^{11} \, \mathrm{cm}^{-2}$, at  $\nu = 4.68$ the density of charge in the 68\% filled lower spin branch TLL is $n_{L} = 5.1\times 10^{10}\, \mathrm{cm}^{-2}$ and as a consequence the sliding CDW would carry  $9\%$ of the charges present in this LL. The observation of narrow band noise at biases above the depinning voltage, and given that sliding CDWs are generally observed with depinning of 1D electron crystals, suggests here the occurrence of two depinning transitions. First, a depinning transition occurring in the easy direction allowing for the current to flow, then a sliding of the crystal, likely in the perpendicular direction. We conclude that an electron nematic liquid crystal which is susceptible of being pinned in both directions is a good candidate to support our observations, and consistent with the previous work of Ref.\cite{N2Yacobi}.

\begin{figure}[!h]
\includegraphics[width=9cm]{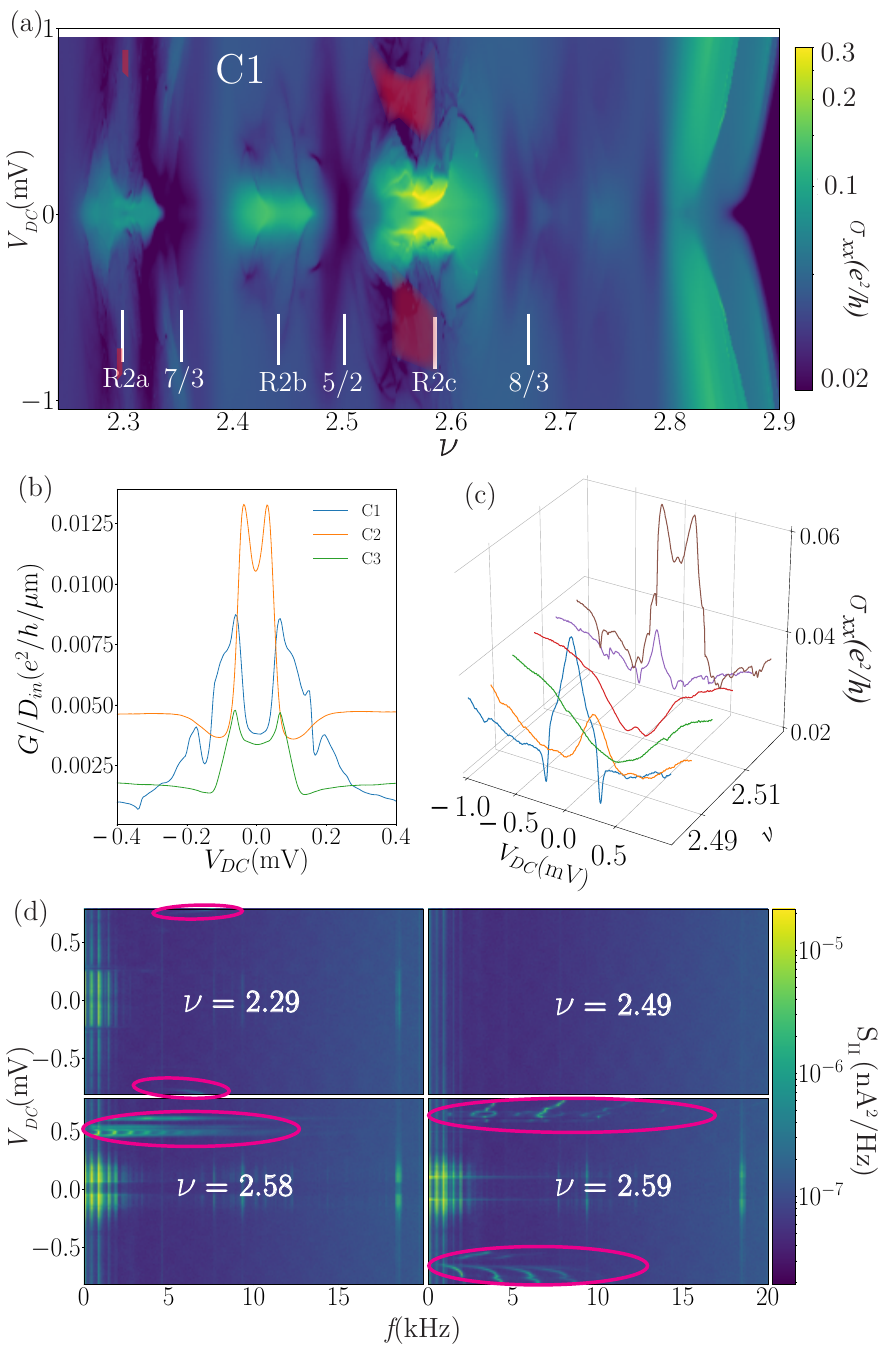}
\caption{ (a): $\sigma_{xx}$ in the SLL \textit{vs.} $V_{DC}$ and $\nu$ in C1. Narrow band noise was observed in red overlay regions. (b) $G/D_{in}$ \textit{vs.} $V_{DC}$ in R2c center ($\nu=2.58$) for the three samples. (c) $\sigma_{xx}$ \textit{vs.} $V_{DC}$ at different filling factors around $\nu=5/2$ in C1. (d) Current noise spectral density $S_{II}(f)$ \textit{vs.} frequency $f$ at $\nu = 2.29,\, 2.49,\, 2.58,\, 2.59$.}
\label{QHE2}
\end{figure}

Similar measurements have been carried out in the SLL and reported in Fig. \ref{QHE2}. The Corbino geometry does not provide any information regarding the Hall resistance and as such it is difficult to unambiguously identify the RIQHE. However, the position of the conductance maxima as well as their temperature dependence (see for example Fig. \ref{C1carac}(b) for C1) can be compared and identified with previously RIQHE states and identified as R2a, R2b, R2c, per the nomenclature of Refs. \cite{reentrant1,reentrantT,GaborR}. Focusing on R2c in C1 (Fig. \ref{QHE2}(a)), we observe that it holds peculiar similarities with the stripe phases observed in the TLL. First, the maximum conductivity $\sigma_{xx}^{max}$ is singularly enhanced at low temperatures and reaches $\sigma_{xx}^{max-20mK}=0.3\, e^2/h$ with $\sigma_{xx}^{max-20mK}/\sigma_{xx}^{max-200mK}>10$ (Fig. \ref{C1carac}(b)). Comparing it to the CDWs in the TLL, such an increase is observed only in the stripe phase where at $\nu = 9/2$ $\sigma_{xx}^{max-20mK}/\sigma_{xx}^{max-200mK}\simeq 20$. This behavior is {\it not} observed in the bubble phases. These peaks in conductance are present in R2a, R2b and R2c in all our Corbino samples and while they resemble to the stripe conductivity maxima observed  at $\nu = 9/2$, our data however show that they disappear (or melt) at lower Joule power $P$ dissipated in the sample. For example, $P \simeq 3\, \mathrm{pW}$ was sufficient to destroy the conductance peak in R2c whereas a $P \simeq 18\, \mathrm{pW}$ is required to destroy the stripe phase in C1 at $\nu = 9/2$. We take this as evidence disfavoring scenarios based on isotropic bubble phases because it is difficult to reconcile it with the conductance enhancement at low temperatures for these RIQHE states. However, these phases can not be associated with the stripe phase occurring  at $\nu = 9/2$ because here the conductance is not proportional to the ring diameter $D_{in}$ (Fig. \ref{QHE2}(b)).

Our $\sigma_{xx}$ data also suggest a coexistence occurring between CDWs and FQHE states in the SLL as was previously reported in \cite{phaseComp2} with optical measurement. To illustrate it, we focus on the $\nu = 5/2$ FQHE state and its vicinity (Fig. \ref{QHE2}(c)). At $\nu = 5/2$, the conductivity minima is at zero bias voltage, as expected for a FQHE state (green curve in Fig. \ref{QHE2}(c)). As soon as we depart from the exact filling factor $\nu = 5/2$, a maximum in the conductivity is observed at zero bias voltage followed by a local minimum at finite bias (red and orange  curves in Fig. \ref{QHE2}(c)). Already at $\nu=2.52$ (brown curve in Fig. \ref{QHE2}(c)) the conductivity resembles that of the RIQHE phase, i.e. it decreases with increasing bias voltage as crystalline order melts, as in Fig. \ref{QHE2}(b). Previous work suggested the onset melting energy for RIQHE states to be lower than the FQHE gap \cite{GaborR, 52gap}, and since $\sigma_{xx}$ give a direct access to bulk conductivity without any contribution from edge states (unlike in \cite{Miller2007, 5/2tunneling, 5/2tunneling2}),  the zero bias peak in the conductivity points towards phase coexistence.  First, the crystal melts with increasing bias voltage whereas the FQHE state vanishes at higher bias voltage.  A similar situation where a competition occurs between a series of FQHE states and the Wigner crystal, albeit with increasing temperature rather than DC bias, has been previously reported deep in the first Landau level\cite{CDWFQHE}. We also observe a similar coexistence between electron liquids and solids at filling factors in the vicinity of $\nu = 7/3$ and $\nu = 8/3$ FQHE states, although in a less pronounced way.

NBN was also observed in the SLL for R2c and to a lesser extent in R2a, as shown in Fig. \ref{QHE2}(d). This is to our knowledge the first observation of NBN in the SLL. Here, the typical fundamental frequency observed is five to ten times lower than in the TLL. Applying a similar interpolation technique we find in R2c (at $\nu = 2.58$) that $I_{CDW}/f_0  \approx 8.43\times 10^{-14}\, \mathrm{A\,s}$ \cite{suma}. Given that here the cyclotron length is $l_{c} = 23.4 \, \mathrm{nm}$, we estimate the charge density in the sliding CDW to be  $n_c \simeq 2.2 \times 10^{9} \, \mathrm{cm}^{-2}$ with a drift velocity of $v_d = 23.4\, \mu \mathrm{m.s^{-1}}$. At $\nu = 2.58$, given the electron density in the last LL being  $n_{L} =  5.52 \times 10^{10}\, \mathrm{cm}^{-2}$, the sliding CDW corresponds to $4\%$ of the SLL lower spin branch electron density. This is consistent with what we have observed in the TLL and provides confidence that a sliding CDW is a reasonable explanation. Finally,  as in the TLL the sliding CDW points towards a depinning transition of an electron liquid crystal implying here that such a phase is present in re-entrant states R2a and R2c.

To conclude, narrow band noise observations show that CDWs in the TLL and RIQHE states R2a and R2c in the SLL very likely contain a stripe-like order susceptible of being pinned such as for a nematic electron liquid crystal. The combination of these observations with conductivity maps {\it versus} DC bias taken in the SLL and TLL show that RIQHE states R2a, R2b and R2c are unlikely to be bubble states, and instead points toward an electron liquid crystal ordering which can coexist with FQHE liquids.

This work has been supported by Canada Excellence Research Chairs program, NSERC, CIFAR, MDEIE, FRQNT {\it via} INTRIQ and the Canada
Foundation for Innovation. The work at Princeton University was funded by the Gordon and Betty Moore Foundation through the EPiQS initiative Grant GBMF4420, and by the National Science Foundation MRSEC Grant DMR-1420541. The authors acknowledge fruitful discussions with Ren\'{e} C\^{o}t\'{e}, Edouard Pinsolle, and Ganesh Sundaram as well as technical help of Gabriel Lalibert\'{e} and Oulin Yu.

\section*{Supplementary material for: "Competing charge density waves probed by non-linear transport and noise in the second and third Landau levels" }

\subsection*{Sample Fabrication}
Contacts to the 2D electron gas were made with the deposition of  26/54/14/100 nm of Ge/Au/Ni/Au after a UV lithography made with a photoplotter. In a following step, an annealing was made in a rapid thermal annealer furnace with a two plateaus recipe 370/450$^{\circ}$C for 30/60 seconds in an $\mathrm H_2\mathrm N_2$ atmosphere. A blue LED was used to illuminate the sample while cooling down to $\sim 4\mathrm{K}$.

\begin{figure}[!h]
\begin{center}
\includegraphics[width=9cm]{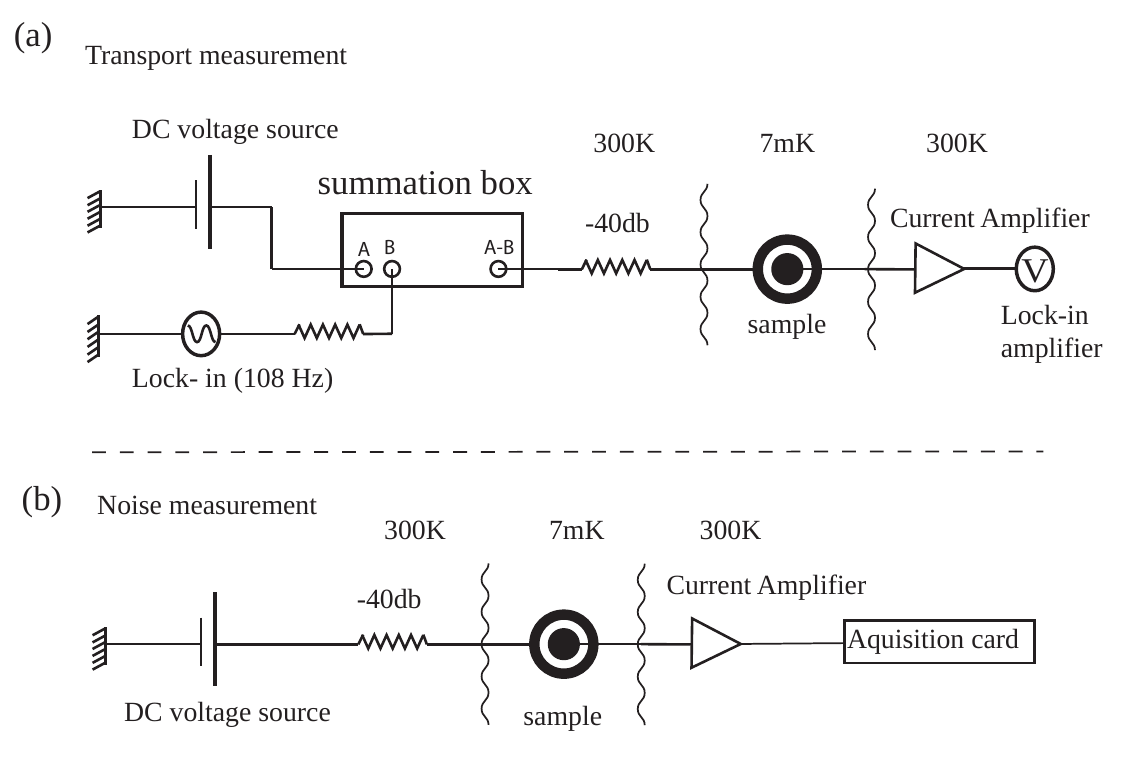}
\end{center}
\caption{{\small Schematics of differential conductance (a) and noise (b) measurements.}}
\label{schema}
\end{figure}

\subsection*{Experimental Scheme}
The sample was placed inside a silver box thermally connected to the mixing chamber with a 2 mm silver wire. Fig. \ref{schema} show schematics for the transport and noise measurements. The differential conductance of the samples was measured using an AC voltage excitation ($10\mu \mathrm{V}$ at 108 Hz) superimposed on a DC bias that could be swept while the AC current was recorded using a current pre-amplifier and a lock-in amplifier. Noise was measured at low frequency using a current amplifier (NF - CA5350) and an acquisition card. The signal from the amplifier was recorded during 1s with a sampling rate of $102.3 \mathrm{k}$ per second. Each point was measured 30 times and then averaged.

\begin{figure}[!h]
\begin{center}
\includegraphics[width=8cm]{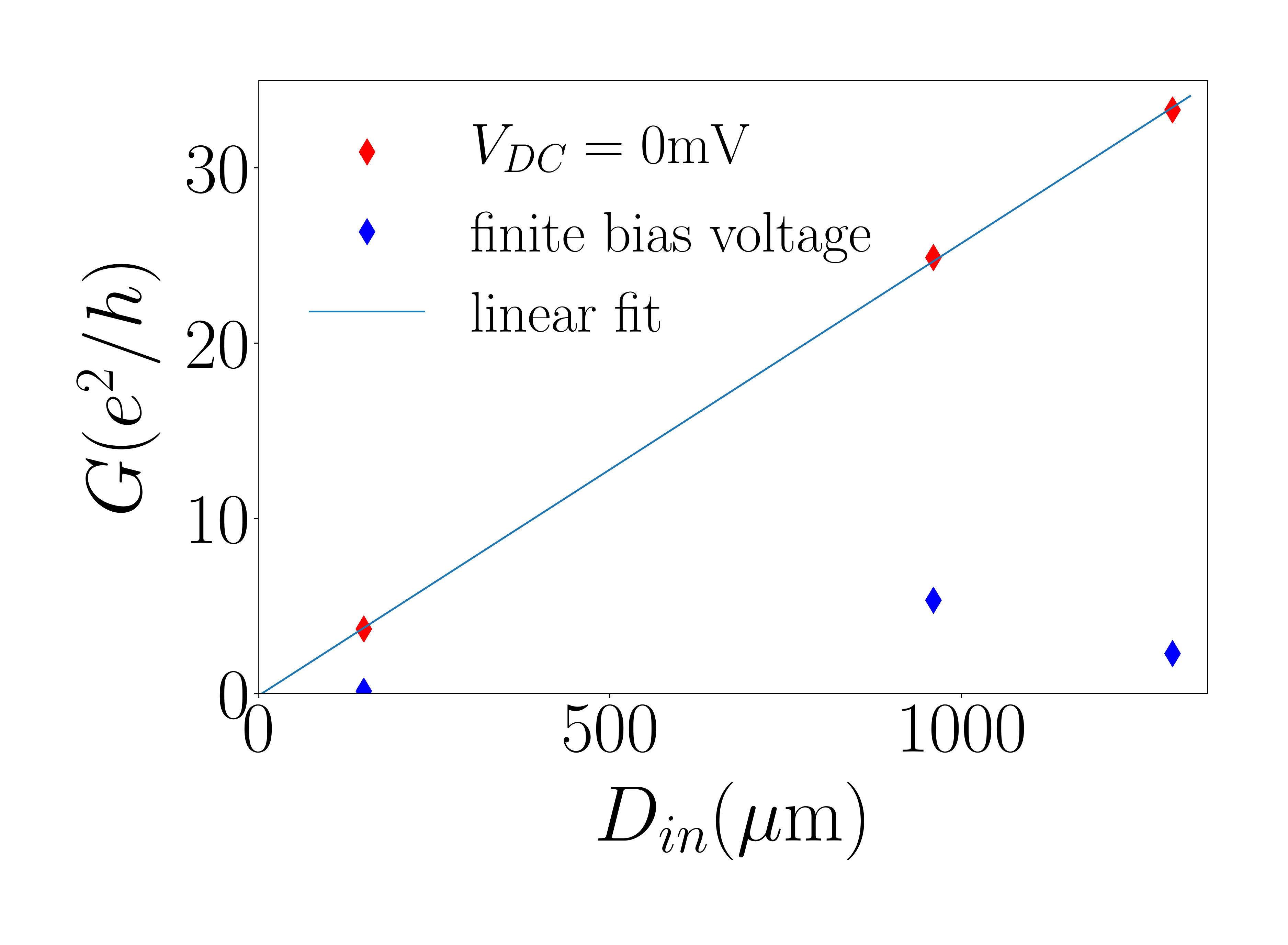}
\end{center}
\vspace*{-7mm}
\caption{{\small Conductance $G$ \textit{vs.} inner ring diameter $D_{in}$ of the different Corbino samples at $V_{DC} = 0 $ (red losanges)  and at bias voltage where the stripe phase has completely melted (blue losanges): $V_{DC} = -2.7$ mV for C1,$V_{DC} = -1.2$ mV for C2, $V_{DC} = -1.8$ mV for C3. The blue line is a linear fit to the red losange data points. }}
\label{GvsD}
\end{figure}

\subsection*{Additional Measurements}

Fig. \ref{GvsD} shows conductance \textit{versus} inner contact diameter $D_{in}$ in the three samples at $V_{DC} = 0 $ (red losange), as well as at a bias voltage where the stripe phase is melted (blue losange). The linear fit of the conductance at zero bias yields $G = (0.026 \times D_{in} - 0.12)\, e^2/h$. \\\

\begin{figure}[!h]
\begin{center}
\includegraphics[width=9cm]{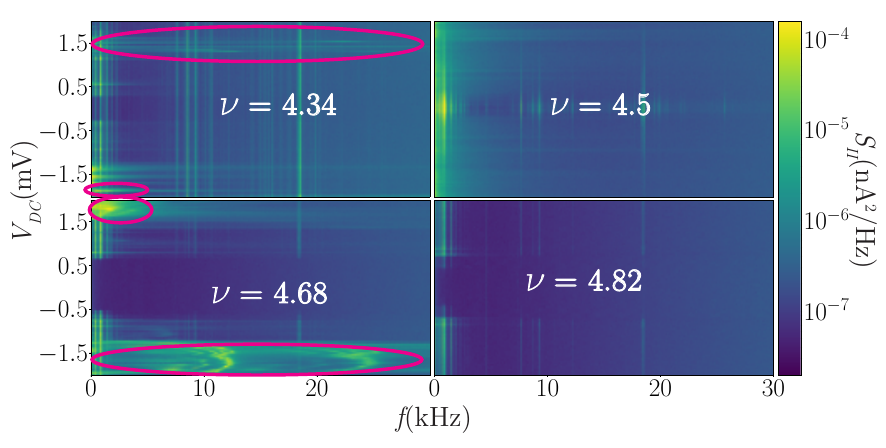}
\end{center}
\vspace*{-4mm}
\caption{Current noise spectral density $S_{II}$  \textit{vs.} frequency $f$ at $\nu = 4.34$,\, 4.5,\, 4.68,\ and 4.82.}
\label{ContourTLL}
\end{figure}

Fig. \ref{ContourTLL} shows color intensity plots of the noise \textit{versus} frequency and bias voltage at filling factors $\nu = 4.34,\, 4.5,\, 4.68$ and  4.82.\\

\begin{figure}[!h]
\begin{center}
\includegraphics[width=9cm]{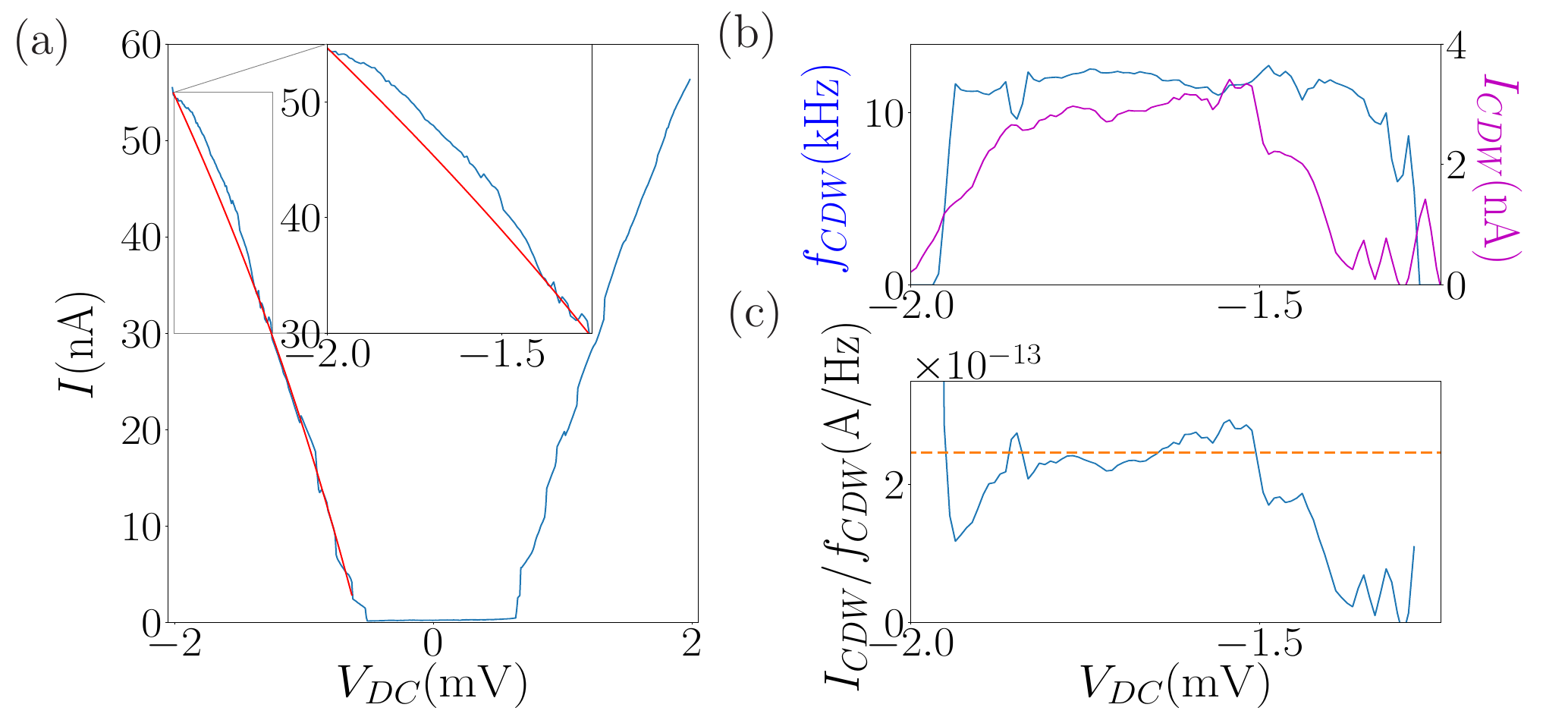}
\end{center}
\vspace*{-2mm}
\caption{(a) The blue curve is the DC current $I(V)$ measured \textit{vs.} DC bias voltage $V_{\mathrm{DC}}$ for sample C1 at $\nu = 4.68 $. The red curve is a cubic polynomial fit of the $I(V_{DC})$ curve. Inset: close-up on the narrow band noise region. (b) The blue curve is the fundamental frequency $f_{CDW}$ of the narrow band noise \textit{vs.} $V_{\mathrm{DC}}$, the purple curve is the excess current flowing $I_{CDW}$ \textit{vs.} $V_{\mathrm{DC}}$ in the NBN region. It is obtained from the subtraction of the polynomial fit to $I(V)$. (c) Ratio of $I_{CDW}$ to $f_{CDW}$ \textit{vs.} $V_{\mathrm{DC}}$.}
\label{f1}
\end{figure}

\begin{figure}[!h]
\begin{center}
\includegraphics[width=9cm]{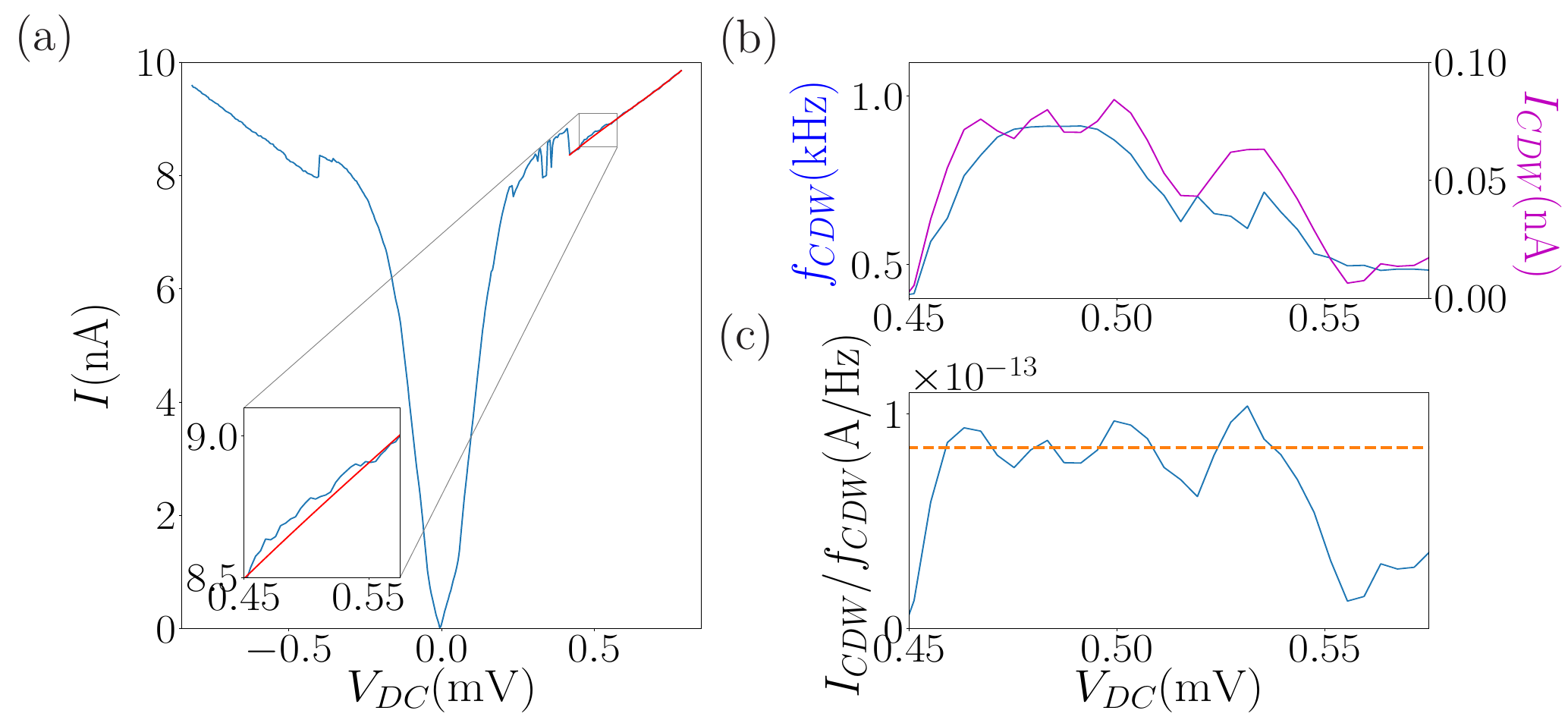}
\end{center}
\caption{(a) The blue curve is the DC current $I(V_{DC})$ measured \textit{vs.} DC bias voltage $V_{\mathrm{DC}}$ for sample C1 at $\nu = 2.58 $. The Red curve is a cubic polynomial fit of the $I(V_{DC})$ curve. Inset: close-up on the narrow band noise region. (b) The blue curve is the fundamental frequency $f_{CDW}$ of the narrow band noise \textit{vs.} $V_{\mathrm{DC}}$, the purple curve is the excess current flowing $I_{CDW}$ \textit{vs.} $V_{\mathrm{DC}}$ in the NBN region. It is obtained from the subtraction of the polynomial fit to $I(V_{DC})$. (c) Ratio of $I_{CDW}$ to $f_{CDW}$ \textit{vs.} $V_{\mathrm{DC}}$.}
\label{f2}
\end{figure}

Fig. \ref{f1} and fig. \ref{f2} show how the current carried by sliding CDW ($I_{CDW}$) was extracted from the $I-V_{DC}$ curves at $\nu = 4.68$ in the third Landau level (TLL), and at $\nu = 2.58$ in the second Landau level (SLL). Fig. \ref{f1}c and fig. \ref{f2}c  clearly demonstrate that the excess current $I_{CDW}$ extracted from the $I(V_{dc})$ characteristics is proportional to the fundamental frequency of the observed NBN.    \\

\begin{figure}[!h]
\begin{center}
\includegraphics[width=9cm]{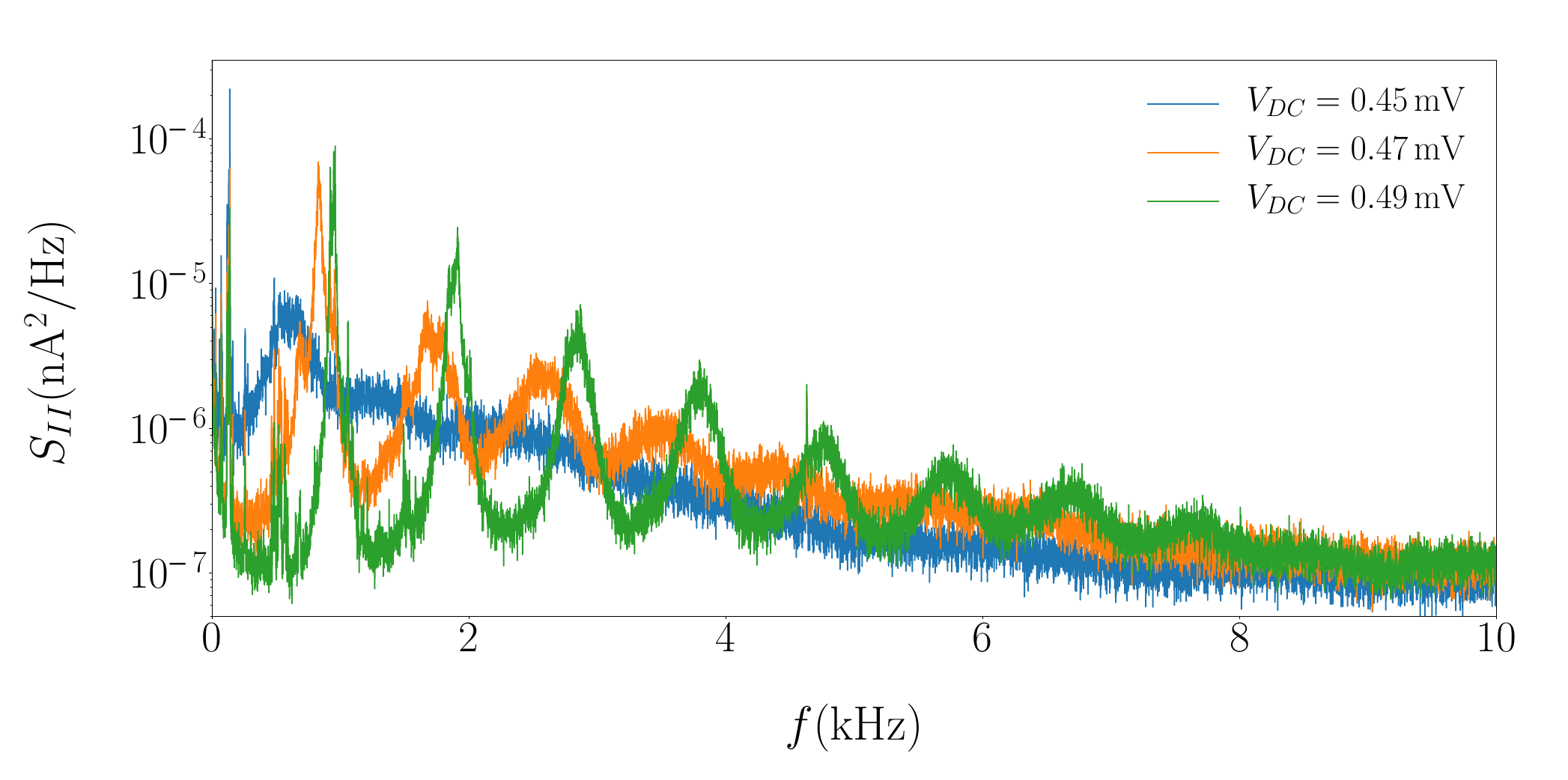}
\end{center}
\vspace*{-2mm}
\caption{Current noise spectral density $S_{II}$  \textit{vs.} frequency at three different $V_{\mathrm{DC}}$ in the SLL at  $\nu = 2.58 $.}
\label{f3}
\end{figure}

Fig. \ref{f3} shows an example of narrow band noise at three different bias voltages in the SLL at  $\nu = 2.58 $.\\

\begin{figure}[!h]
\begin{center}
\includegraphics[width=9cm]{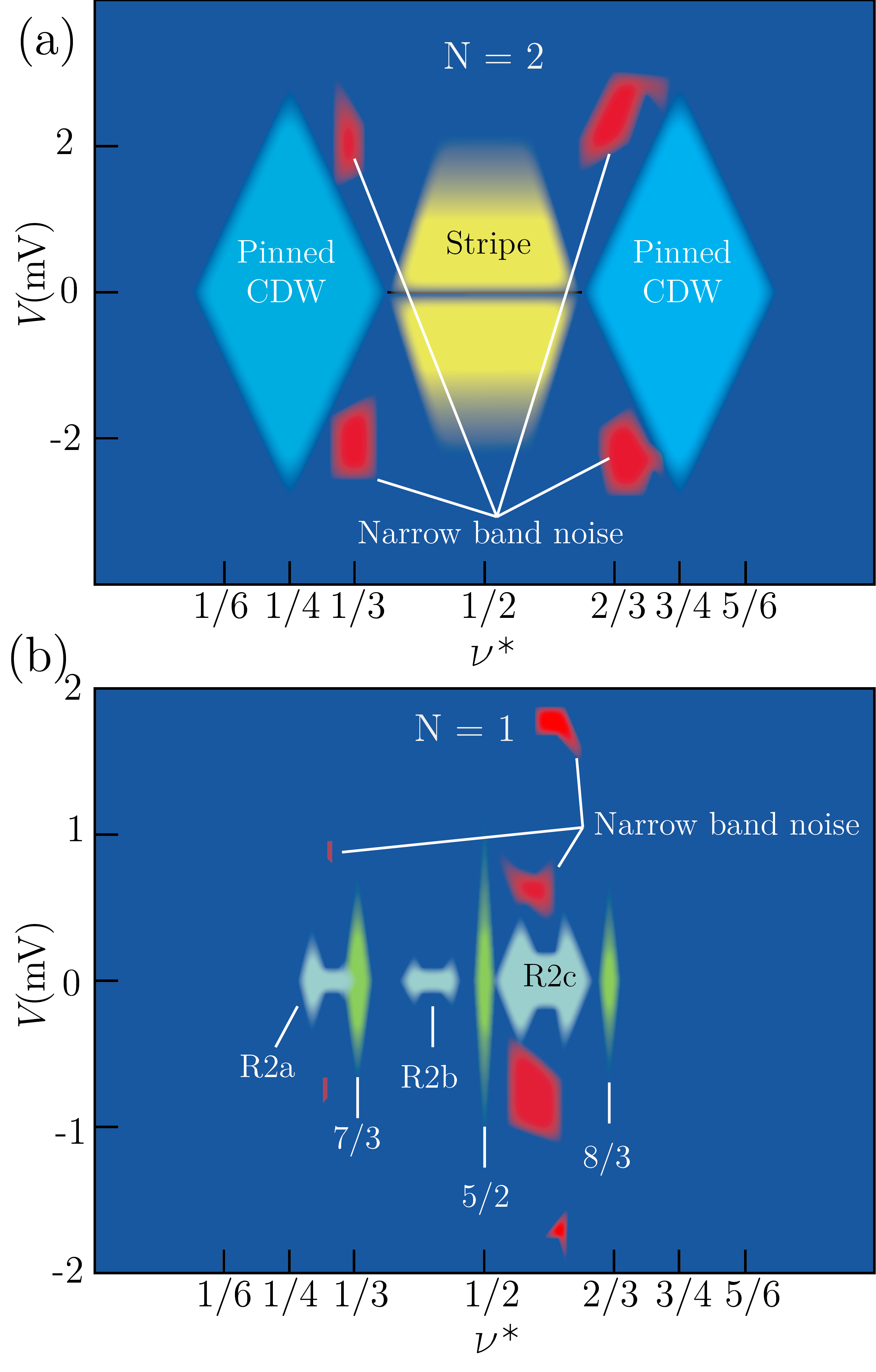}
\end{center}
\caption{ Map of states observed in the TLL (a) and in the SLL (b). These states are shown \textit{vs.} partial filling factor $\nu^*$ and $V_{\mathrm{DC}}$. In the SLL, $7/3$, $5/2$, $8/3$ are fractional quantum Hall states and R2a, R2b and R2c are re-entrant integer quantum Hall states.}
\label{f4}
\end{figure}

Fig. \ref{f4} shows our deduced map of the different electronic states observed in the TLL and SLL.

\end{document}